\documentclass[9pt,twocolumn,twoside]{osajnl}

\journal{ol} 

\setboolean{shortarticle}{true}

\usepackage{makecell}

\title{Stable tuning of photorefractive micro-cavities using an auxiliary laser}

\author[1,$\dagger$]{Joshua B. Surya}
\author[1,$\dagger$]{Juanjuan Lu}
\author[1]{Yuntao Xu}
\author[1,*]{Hong X. Tang}

\affil[1]{Department of Electrical Engineering, Yale University, New Haven, Connecticut 06511, USA}
\affil[$\dagger$]{These authors contributed equally to this letter}
\affil[*]{Corresponding author: hong.tang@yale.edu}

\begin{abstract}
Cavity nonlinear optics enables intriguing physical phenomena to occur at micro- or nano-scales with modest input powers. While this enhances capabilities in applications such as comb generation, frequency conversion and quantum optics, undesired nonlinear effects including photorefraction and thermal bistability are exacerbated.  In this letter, we propose and demonstrate a highly effective method of achieving cavity stabilization using an auxiliary laser for controlling photorefraction in a z-cut periodically poled lithium niobate (LN) microcavity system.  Our numerical study accurately models the photorefractive effect under high input powers, guiding future analyses and development of LN microcavity systems.  
\end{abstract}

\setboolean{displaycopyright}{true}

\begin{document}

\maketitle


Over the past few years, a surge in development of chip-scale applications on thin-film lithium niobate ($\mathrm{LiNbO_3}$, LN) has propelled this material platform into becoming one of the top contenders for the next generation of nanophotonic materials.  Achievements such as wide-band, low-threshold electro-optic (EO) combs \cite{Zhang:broadband_eocomb}, dissipative Kerr solitons \cite{He:soliton, Gong:2umsoliton}, microwave-to-optical converters \cite{Shao:s,youssefi2020cryogenic}, as well as ultra-high efficiency second harmonic generators (SH, SHG) \cite{jjlu:250k,rluo:tunable_shg,Chen:pplnring_shg,Wang:waveguide_shg}, among other applications \cite{Shao:s,Okawachi:20,ayed_singlephoton,jjlu:1percent,rluo:OPG,Jankowski:20,mYu:two_octave,sWang2020:erbium_ln}, are proof of the vast potential of LN.  Invariably, most of the current research and progresses on LN have focused on its strengths, i.e. the material's large $\chi^{(2)}$ and $\chi^{(3)}$ nonlinearities \cite{fejer:qpmshg,Gong:2umsoliton,Wang:waveguide_shg,mYu:two_octave,Weis1985:summary}, while relatively few studies have addressed a commonly observed issue regarding the stability of LN microresonators \cite{xSun:oscillation_dynamics,hJiang17_fastresponse,jWang16_photothermaleffects,aSavchenkov2006:enhancement_PR}.  Namely, photorefractive (PR) effect \cite{Villarroel10_prdamage,Yariv96_holographicmemory}, an EO effect arising from the static field-induced refractive index change due to a redistribution of space-charge caused by the excitation of trap ions, which is significantly amplified in systems with high optical quality (Q) factors.  

Recent works on the PR effect of nanophotonic LN microresonators have highlighted the difficulty in maintaining precise control over the cavity wavelength due to strong tuning and hysteretic effects of photorefraction \cite{jjlu:250k,xSun:oscillation_dynamics,Holzgrafe:20}. With some previous studies indicating that additional processing can amplify the PR effect in microresonators by injecting additional trap states \cite{yYang03_opticalstorage}. The instability is further complicated by the fact that recombination times of the generated charge carriers due to PR effect can be unpredictable, ranging from milliseconds to minutes at room temperature \cite{jjlu:250k,Yariv96_holographicmemory,mLuennemann03:improvements_of_indexchanges,hJiang17_fastresponse,xSun:oscillation_dynamics}.  Current mitigation techniques typically rely on utilizing low input powers, operating in high environment temperatures, and minimal post-processing. These efforts have limited effectiveness which do not provide long-term, universal solutions.  

Another approach, first suggested in a previous work by Sun et al. \cite{xSun:oscillation_dynamics}, aimed at quenching the PR effect in a cavity system by repeatedly scanning over a cavity mode, gradually increasing the space-charge electric field within the resonator. This method successfully reduces instability and elegantly shows that saturation of the PR effect can be achieved in microresonators.  Nevertheless, an effective and targeted method allowing flexible and stable control over cavity wavelengths is yet to be demonstrated. Such a solution can significantly alleviate current challenges in the LN integrated nanophotonics community.  

In this letter, a method of achieving stable wavelength tuning by using an auxiliary laser drive is described. The effectiveness of this scheme is demonstrated in our z-cut periodically-poled lithium niobate (PPLN) microcavity system, where a phase-matched SHG output was tuned and stabilized. The saturation behavior of the PR effect in such system was described by our generalized three-mode coupled equations, and the results of our measurements were captured accurately by the model. We further show from our tuning measurements as well as analytical and numerical studies of the multi-modal system, that depending on power requirements, the cavity wavelength is highly manipulable. The presented work can be applied to a range of microcavity systems where PR instability may be significant.

\begin{figure}[t]
\centering
\includegraphics[width=\linewidth]{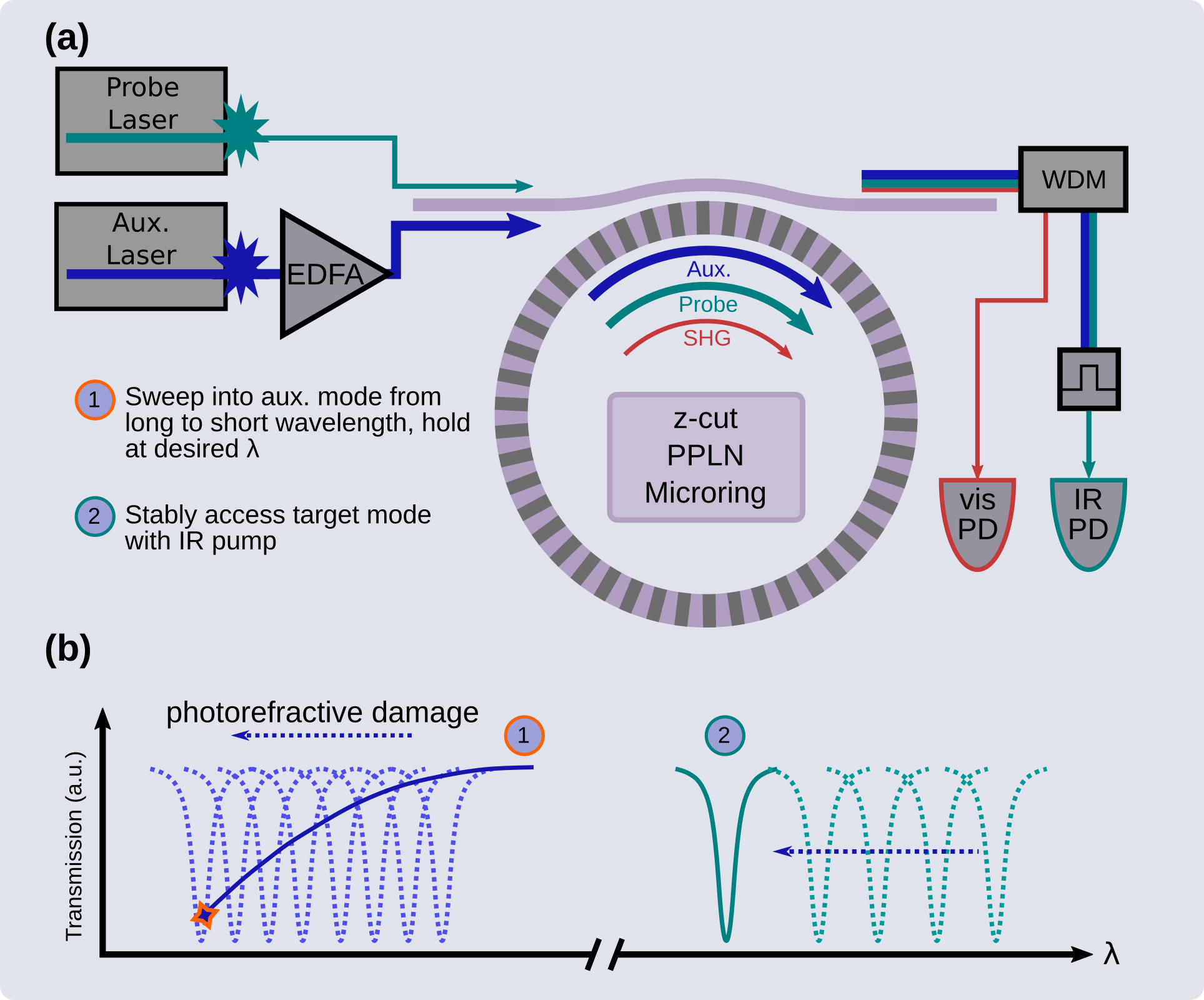}
\caption{(a) Schematic of the experimental setup. Two laser inputs, a pump probe and an auxiliary drive were used for interrogating two separate cavity modes (the pump mode $a$ and auxiliary mode $c$). A higher power for the auxiliary laser input is used. (b) Conceptual diagram of the operation principle. The photorefractive effect causes unstable and difficult resonance locking. The proposed solution first utilizes an auxiliary laser, by slowly tune it into resonance at high power, where stable tuning is possible due to the stable dynamics of thermal expansion and photorefraction. With the auxiliary laser
in resonance, the target mode can be stably accessed at low powers. }
\label{fig:setup}
\end{figure}

The analysis of our system begins by introducing the simplified three-mode interaction Hamiltonian of the LN microring cavity under a unitary transformation in the reference frame of the pump and auxiliary drive laser,
\begin{equation}
    \begin{alignedat}{1}H= & \ \delta_{a}a^{\dagger}a+\delta_{b}b^{\dagger}b+\delta_{c}c^{\dagger}c+g\left(a^{2}b^{\dagger}+\left(a^{\dagger}\right)^{2}b\right)\\
    + & \ i\epsilon_a\left(-a+a^{\dagger}\right) + i\epsilon_c\left(-c+c^{\dagger}\right)
    \end{alignedat}
\label{eq:simple_hamiltonian}
\end{equation}
where $a$, $b$, and $c$ respectively denote the pump, SH, and auxiliary modes, normalized such that $\left|a\right|^2$, $\left|b\right|^2$, and $\left|c\right|^2$ represent the number of photons in each mode. $\delta_{a(b,c)}=\omega_{a(b,c)}-\omega_{fa(b,c)}$ is the detuning from the pump telecom cavity mode (SH cavity mode, telecom drive mode), and $g$ is the nonlinear coupling rate of the SH process. The pump (drive) term is represented by $\epsilon_{a(c)}=\sqrt{2\kappa_{a(c),1}\frac{P_{a(c)}}{\hbar\omega_{fa(c)}}}$. In our work, the auxiliary drive mode was selected to be TE in order avoid phase-matching with any particular $\chi^{\mathrm{(2)}}$ process that would interfere with SHG and is designated for the sole purpose of PR effect saturation.

In addition to significant $\chi^{(2)}$ nonlinear optical effects, under high input powers, thermal and PR interactions intensify as the input wavelength approaches resonance. The thermal shift of the cavity resonance is due to two underlying effects; the thermo-optic effect which influences the refractive index, as well as the thermal expansion of the microcavity \cite{Carmon2004:thermal_behavior,Surya:18}. In comparison, the PR effect is an alteration of the refractive index by the EO effect, induced by a static space-charge electric field $E_{sc}$ due to the drift and diffusion of carriers generated by the photoionization of trap states \cite{kukhtarev1976:photorefraction}. This refractive index change counteracts the thermo-optic effect, and is represented by $\Delta{n}=-\frac{n_{\mathrm{eff}}^3}{2}r_{\mathrm{eff}}E_{\mathrm{sc}}$. Where $n_{\mathrm{eff}}$ is the effective index of the resonator and $r_{\mathrm{eff}}$ is the effective EO coefficient.

Bulk LN is known to possess numerous trap states. Furthermore, there is an increased likelihood in the ionization of trap donors due to the optical confinement as well as the large enhancement of intracavity power when the input laser wavelength approaches resonance. Driven by LN's intrinsically large EO tensor coefficients, the circulating intracavity power induces a substantial resonance wavelength shift even at modest input powers. 

Fig.\,\ref{fig:setup}a presents a high-level schematic of the experimental setup, where an auxiliary laser is first slowly tuned into the cavity resonance at a high input power, followed by the input of a probe laser used to stably access a target mode with efficient SHG. Conceptually, this is illustrated in Fig.\,\ref{fig:setup}(b), where the PR effect due to an auxiliary laser drive causes a large blue-shift in the resonance wavelengths of the modes, after which by using a low-power probe laser, the resonance of a mode of interest can be stably interrogated. Such scheme makes it possible to access stable SHG power without the use of complex servos.

Under the Heisenberg representation of $a$, $b$, $c$, the coupled nonlinear, thermal as well as space-charge field interactions of the modes are guided by the following dynamic equations,
\begin{equation}
\label{eq:eom}
    \begin{aligned}
        \frac{d}{dt}a= & -\left(i\delta_{a}+\kappa_{a}\right)a-i\left(g_{E}E_{\mathrm{sc}}+g_{T}\Delta{T}\right)a-i2ga^{\dagger}b+\epsilon_{a}\\
        \frac{d}{dt}b= & -\left(i\delta_{b}+\kappa_{b}\right)b-i\left(g_{E}E_{\mathrm{sc}}+g_{T}\Delta{T}\right)b-iga^2\\
        \frac{d}{dt}c= & -\left(i\delta_{c}+\kappa_{c}\right)c-i\left(g_{E}E_{\mathrm{sc}}+g_{T}\Delta{T}\right)c+\epsilon_{c}
    \end{aligned}
\end{equation}
where $\kappa_{a,b,c}$ is the total coupling rate of the cavity modes due to absorption of the material, scattering losses as well as external coupling.  The electro-optic coupling rate is represented by $g_{E}$, whereas the thermal shift coefficient considering both the thermo-optic effect and thermal expansion is given by $g_{T}$. $E_{sc}$ and $\Delta{T}$ denote the space-charge electric field and variation in environment temperature due to the total intracavity photons. 

The dynamics of modes $a,b,c$ shown in Eq.$\thinspace$\ref{eq:eom} are dictated by a second order $\chi^{(2)}$ nonlinear frequency conversion process (SHG) as well as the thermal and PR effects. Thermal dynamic contributions to cavity resonances have been thoroughly analyzed in previous studies \cite{Carmon2004:thermal_behavior,Schmidt2008:thermal_effects,Hu2020:all_optical}, and will not be discussed in detail here. While similar treatments have been applied to electro-optic cavity systems with significant PR effect \cite{xSun:oscillation_dynamics,He:soliton,hJiang17_fastresponse}, care must be taken when modeling such a system under high input powers, since saturation of the charge carrier generation rate could reasonably occur in this regime. In previous studies, it was observed that the space-charge field generation rate decreased with increased input powers \cite{xSun:oscillation_dynamics}.  We address this phenomenon by taking into account the band transport model \cite{kukhtarev1976:photorefraction}, where the probability of trap ionization under illumination is taken into consideration. Consequently, the rate equations of $E_{sc}$ and $\Delta{T}$ can be written as, 
\begin{subequations}
\label{eq:bistability}
    \begin{align}
        \label{eq:photorefraction}
        \frac{dE_{\mathrm{sc}}}{dt}=&
        \begin{cases}
            -\gamma_{e}E_{\mathrm{sc}}+          K_{e}\left(1-\frac{N_{tot}}{N_{\mathrm{eff}}}\right)N_{tot} & N_{tot}\leq\frac{N_{\mathrm{eff}}}{2}\\
            -\gamma_{e}E_{\mathrm{sc}}+\frac{K_{e}N_{\mathrm{eff}}}{4} & N_{tot}>\frac{N_{\mathrm{eff}}}{2}
        \end{cases}\\
        \label{eq:thermal}
        \frac{d\left(\Delta{T}\right)}{dt} &=  -\gamma_{th}\Delta T+K_{th,a}\left|a\right|^2+K_{th,b}\left|b\right|^2+K_{th,c}\left|c\right|^2
    \end{align}
\end{subequations}
where the decay rates are denoted by $\gamma_{e,th}$, while the intrinsic generation coefficients of space-charge field and heat are given by $K_e$ and $K_{th}$ respectively. Here, $N_{tot}=\left|a\right|^2+2\left|b\right|^2+\left|c\right|^2$ is proportional to the total intracavity photon number and corresponds to a mapping of photons to space-charge. In order to describe the decreasing probability of charge carrier generation with increasing donor ionization at higher input powers, an effective field generation coefficient of $K_e\left(1-\frac{N_{tot}}{N_{\mathrm{eff}}}\right)$ is introduced, where $N_{\mathrm{eff}}$ denotes the total number of trap states available for ionization. This subsequently imposes an upper limit on $E_{sc}$. In our numerical models, $N_{\mathrm{eff}}\approx6\times10^8$ provided a good fit to our observations and approximately corresponds to a donor concentration of $N_d\approx5\times10^{17}\,\mathrm{cm^{-3}}$\cite{Yariv96_holographicmemory}.  Other parameters in the numerical model were fitted to be, $\gamma_e=0.25\,\mathrm{Hz}$, $K_e=180\,\mathrm{V/(m\cdot s)}$,  $g_{E,IR}=2.1\times10^4\,\mathrm{m/(V\cdot s)}$, $g_{E,vis}=8.5\times10^4\,\mathrm{m/(V\cdot s)}$, $\gamma_{th}=230\,\mathrm{kHz}$, $K_{th,(a,b,c)}=0.026\,\mathrm{K/s}$, and $g_T=1.68\,\mathrm{(K\cdot s)^{-1}}$. These values are in good agreement with previous LN microring resonator studies \cite{He:soliton,xSun:oscillation_dynamics}.

\begin{figure}[ht]
\centering
\includegraphics[width=0.95\linewidth]{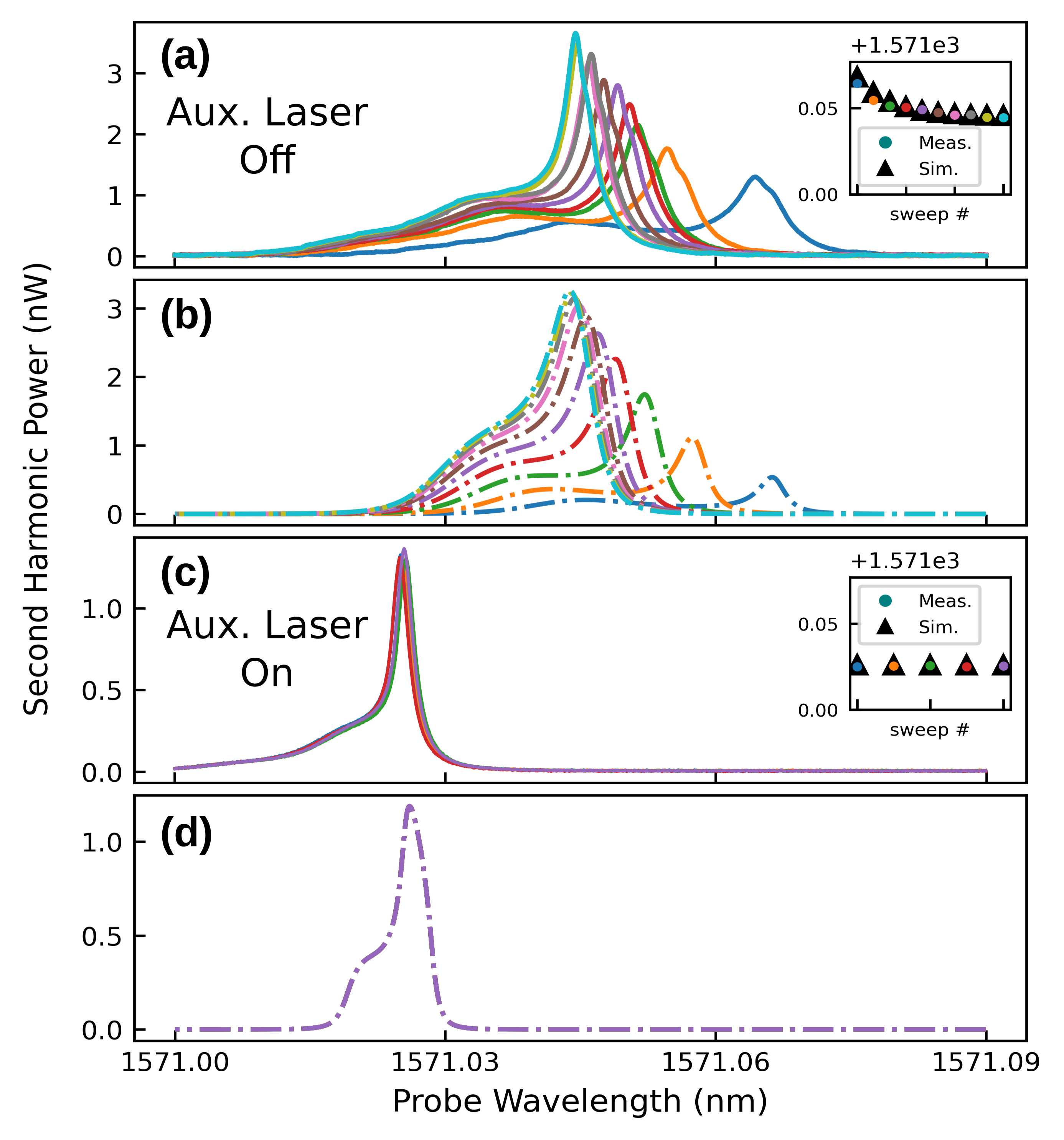}
\caption{Experimental and simulated SH output plots with the auxiliary laser off and on resonance. (a) SH output traces measured at intervals of 10 seconds, with the auxiliary laser set to be far-detuned from the resonance. The on-chip auxiliary and probe laser powers were set to 5.4\,mW and 98.6\,$\mu$W respectively. The inset depicts the shifting peak of the SHG with number of laser scans, where the measured and simulated data are denoted by circles and triangles. (b) The simulated SH output of the measurement without an auxiliary laser.  The cold cavity resonance wavelengths for IR and SH modes are fitted to be approximately 1571.05\,nm and 785.535\,nm respectively, this slight energy mismatch causes the double peaks observed in the SHG spectrum. (c) SH output traces when the auxiliary laser is locked into one resonance, at intervals of 10 seconds. The auxiliary laser suppresses the PR-induced resonance shifts, allowing stable cavity access.  The inset depicts the stabilized peak of the SHG with number of laser scans. (d) Simulated output traces of SHG with the auxiliary laser in resonance.}
\label{fig:multiscan}
\end{figure}

In modeling the system dynamics, we note the assumptions and simplifications made. First, the response times of heat, charge carrier and intracavity photon generation were treated as small compared to the time-scale of the PR effect (i.e. $\frac{d}{dt}a=\frac{d}{dt}b=\frac{d}{dt}c=\frac{d\left(\Delta{T}\right)}{dt}=0$). Second, when $N_{tot}>\frac{N_{\mathrm{eff}}}{2}$, the space-charge field generation rate was maximized such that $K_{e}\left(1-\frac{N_{tot}}{N_{\mathrm{eff}}}\right)N_{tot}=\frac{K_{e}N_{eff}}{4}$, as shown in Eq.\,\ref{eq:photorefraction}. Additionally, every intracavity IR photon in the model had an equal probability in ionizing a trap donor, while visible SH photons were assigned twice that probability. Lastly, for the sake of simplicity, our model only considered the PR response time most relevant to our observations.  

From the analysis of our cavity system above, the effectiveness of this method depends on the condition $\left|c\right|^2\gg\left|a\right|^2$, such that the incremental resonance shift due to probe mode $a$ intracavity photons is sufficiently small.  The latter condition can also be expressed as $g_{E}\Delta{E_{\mathrm{sc}}}\ll\kappa_{a}$, where $\Delta{E_{\mathrm{sc}}}$ is the change in the space-charge field when the probe laser is at resonance.

An important utilization of this method is to help stabilize or fine-tune a cavity resonance for SHG applications. Therefore, to investigate the efficacy of this proposed scheme, we applied the method on a z-cut periodically-poled LN (PPLN) ring resonator and analyzed the measured SHG. The specific device tested has a ring radius of 70\,$\mu$m and a ring width of 1.8\,$\mu$m. The thickness of the LN device is 600\,nm, and is shallow etched with an unetched layer of 180\,nm. A radial poling with a period of $\sim$3$\mu$m is utilized to realize the phase-matching between TM fundamental and TM SH modes using the $\mathrm{d_{33}}$ term of the second order susceptibility tensor. Detailed information about the fabrication and devices can be found in previous works by the authors \cite{jjlu:1percent}. 

To characterize the shift in resonance due to the PR effect, a SH phase-matched resonance was repeatedly scanned over using a Santec TSL710 telecom laser source at an on-chip power of 98.6\,$\mu$W, and the SH output traces were collected.  Initially, the auxiliary laser source was set to be 5.4\,mW (on-chip) at a wavelength off-resonance. Figure\,\ref{fig:multiscan}a plots the ten separate scans measured at a sweeping speed of 0.5\,nm/s, where each scan is initiated at an interval of ten seconds. Overall, a significant blue-shift in resonance wavelength due to the buildup of space-charge field was observed.  Each successive laser scan led to an increase in $E_{sc}$ and less relative blue-shift compared to the previous scan. This is expected since the driving forces of space-charge relaxation is proportional to $E_{sc}$ as described in Eq.\,\ref{eq:photorefraction}. The loaded Q of the IR probe resonance was measured to be $Q_l=1.88\times10^5$. The resonance shifts due to the PR effect is described by our models as depicted in the simulated SHG curves of Fig.\,\ref{fig:multiscan}b. When the auxiliary laser was locked into one resonance, which was selected to be a TE mode at a wavelength of 1554.7\,nm in order to avoid $\chi^{(2)}$ interactions with the probe input (at $\sim$1571\,nm), the buildup of intracavity photons such that $\left|c\right|^2>\left|a\right|^2$ allowed the target mode to be scanned without causing shifts in either the IR or the visible resonance wavelengths. Consequently, with the same sweep rate, subsequent scans no longer caused blue-shifts in the cavity wavelength, as indicated by the measured SHG in Fig.\,\ref{fig:multiscan}c. Furthermore, the peak of the SH output could be stably accessed by hand-tuning. The simulated traces of the SH output with the auxiliary laser in resonance are overlaid in Fig.\,\ref{fig:multiscan}d.

\begin{figure}[h]
\centering
\includegraphics[width=0.95\linewidth]{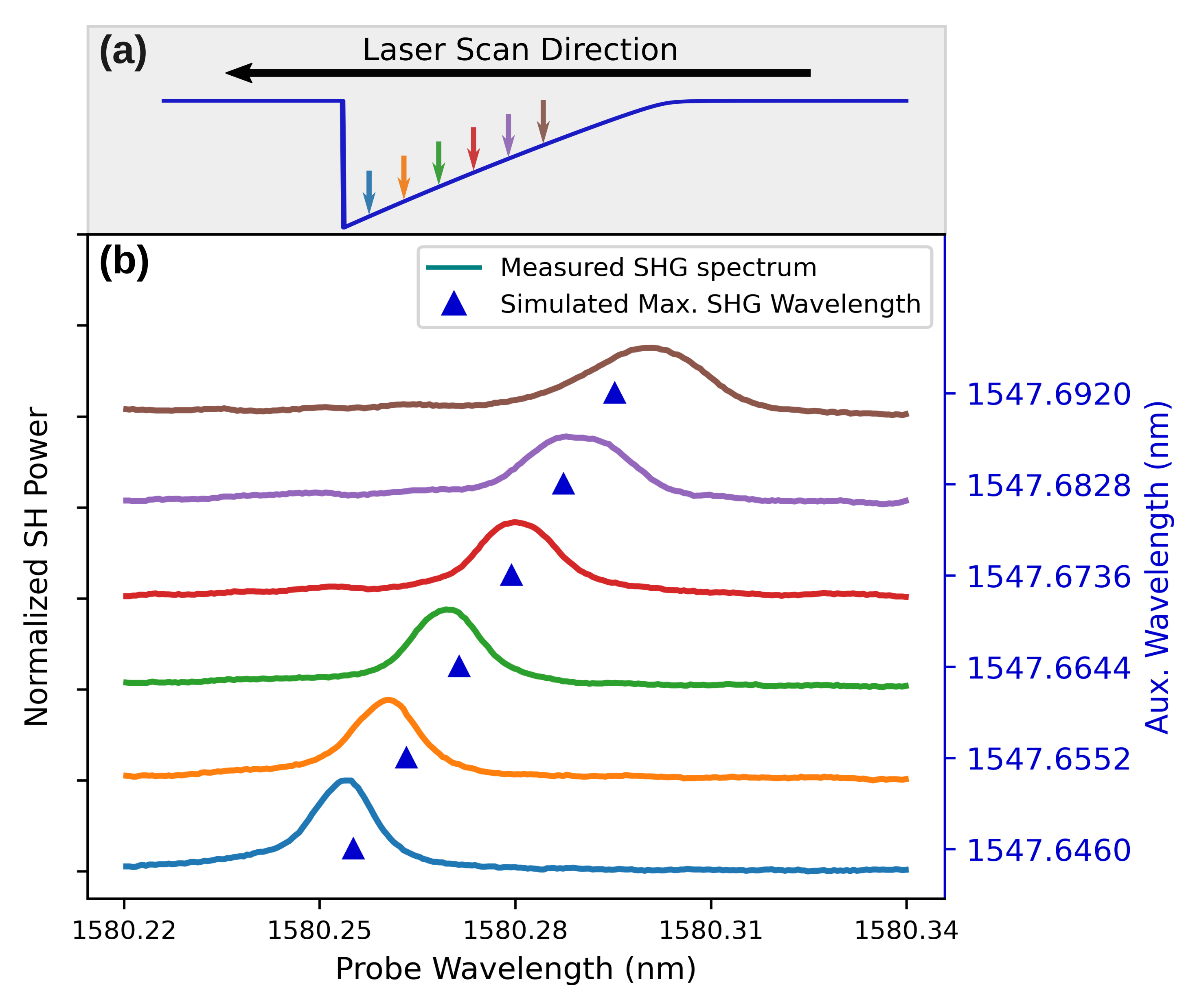}
\caption{Tuning of SHG peak wavelength using an auxiliary laser and the photorefractive effect. (a) Illustration of the transmission curve of the auxiliary mode resonance with red-to-blue scan. Each arrow independently signifies a different auxiliary laser wavelength used when measuring the SHG output. The color of the arrow corresponds to the trace in (b).  (b) Normalized SHG output at various Auxiliary laser wavelengths (5\,mW auxiliary laser power on-chip).  The simulated SHG peak wavelength is plotted against the traces (98.6\,$\mu$W on-chip probe power).}
\label{fig:tuning}
\end{figure}

The resonance wavelength tuning capabilities of this method was further investigated by measuring the SHG of the device at varying auxiliary laser detunings, illustrated in Fig.\,\ref{fig:tuning}a. Here, due to the PR bistability of the microresonator, each detuning corresponded to a different $|c|^2$. This caused varying shifts in the resonance wavelength of the probe and SH mode. The tuning is depicted in the SH output curves of Fig.\,\ref{fig:tuning}b, where each normalized SHG trace corresponds to a different auxiliary laser wavelength shown on the right axis. A change in linewidth of the SHG peaks is observed due to the PR resonance shift being more prominent when $|c|^2$ is low. Additionally, the simulated SH peaks of each auxiliary laser wavelength was plotted against the measured data. Due to environmental perturbations as well as the facet conditions of the photonic chip, fluctuations of the on-chip power with varying auxiliary and probe laser wavelengths can be observed, and are likely the main causes of small discrepancies between the simulated and measured SH peaks. In theory, using higher auxiliary input powers and larger microrings, a wider tuning bandwidth could be achieved. This bandwidth is limited by the competing thermal and PR interactions, the details of which are left for future investigations. 

In this Letter, we have presented a technique for circumventing PR related instabilities in microcavity resonance wavelengths with the assistance of an auxiliary laser. The target probe mode can be reliably and easily accessed even by hand-tuning by saturating the space-charge field. We further utilize this method in the stabilization of a SHG peak in a PPLN microring. The versatility of this scheme is further supported by the extensive numerical modeling of the triple-mode interactions in a dual-resonant, nonlinear system. The theoretical analysis takes into consideration the available trap donor states, which describes the power dependence of the space-charge field generation coefficient. Lastly, the tuning capabilities of this scheme was explored by varying auxiliary laser wavelength detunings from cavity resonance, where picometer precision was achieved. The method described in this Letter can be used in a variety of microresonator systems that suffer from PR instabilities.

\vspace{0.2cm}
\noindent\textbf{Funding.} This work is funded by DOE/BES under award number DE-SC0019406.

\vspace{0.2cm}
\noindent\textbf{Acknowledgements.} The facilities used for device fabrication were supported by the Yale SEAS cleanroom and Yale
Institute for Nanoscience and Quantum Engineering. The
authors thank Kelly Woods, Sean Reinhart, Dr. Yong Sun and Dr. Michael Rooks for
assistance in device fabrication. 

\vspace{0.2cm}
\noindent\textbf{Disclosures.} The authors declare no conflicts of interest.
\bibliography{references}

\begin{thebibliography}{10}
\newcommand{\enquote}[1]{``#1''}

\bibitem{Zhang:broadband_eocomb}
M.~Zhang, B.~Buscaino, C.~Wang, A.~Shams-Ansari, C.~Reimer, R.~Zhu, J.~M. Kahn,
  and M.~Lon{\v{c}}ar, \enquote{Broadband electro-optic frequency comb
  generation in a lithium niobate microring resonator,}
  {\protect\JournalTitle{Nature}} \textbf{568}, 373--377 (2019).

\bibitem{He:soliton}
Y.~He, Q.-F. Yang, J.~Ling, R.~Luo, H.~Liang, M.~Li, B.~Shen, H.~Wang,
  K.~Vahala, and Q.~Lin, \enquote{Self-starting bi-chromatic {LiNb$O_3$}
  soliton microcomb,} {\protect\JournalTitle{Optica}} \textbf{6}, 1138--1144
  (2019).

\bibitem{Gong:2umsoliton}
Z.~Gong, X.~Liu, Y.~Xu, M.~Xu, J.~B. Surya, J.~Lu, A.~Bruch, C.~Zou, and H.~X.
  Tang, \enquote{Soliton microcomb generation at 2\textmu m in z-cut lithium
  niobate microring resonators,} {\protect\JournalTitle{Opt. Lett.}}
  \textbf{44}, 3182--3185 (2019).

\bibitem{Shao:s}
L.~Shao, N.~Sinclair, J.~Leatham, Y.~Hu, M.~Yu, T.~Turpin, D.~Crowe, and
  M.~Lon\v{c}ar, \enquote{Integrated microwave acousto-optic frequency shifter
  on thin-film lithium niobate,} {\protect\JournalTitle{Opt. Express}}
  \textbf{28}, 23728--23738 (2020).

\bibitem{youssefi2020cryogenic}
A.~Youssefi, I.~Shomroni, Y.~J. Joshi, N.~Bernier, A.~Lukashchuk, P.~Uhrich,
  L.~Qiu, and T.~J. Kippenberg, \enquote{Cryogenic electro-optic interconnect
  for superconducting devices,} {\protect\JournalTitle{arXiv preprint
  arXiv:2004.04705}}  (2020).

\bibitem{jjlu:250k}
J.~Lu, J.~B. Surya, X.~Liu, A.~W. Bruch, Z.~Gong, Y.~Xu, and H.~X. Tang,
  \enquote{Periodically poled thin-film lithium niobate microring resonators
  with a second-harmonic generation efficiency of {250,000\%/W},}
  {\protect\JournalTitle{Optica}} \textbf{6}, 1455--1460 (2019).

\bibitem{rluo:tunable_shg}
R.~Luo, Y.~He, H.~Liang, M.~Li, and Q.~Lin, \enquote{Highly tunable efficient
  second-harmonic generation in a lithium niobate nanophotonic waveguide,}
  {\protect\JournalTitle{Optica}} \textbf{5}, 1006--1011 (2018).

\bibitem{Chen:pplnring_shg}
J.-Y. Chen, Z.-H. Ma, Y.~M. Sua, Z.~Li, C.~Tang, and Y.-P. Huang,
  \enquote{Ultra-efficient frequency conversion in quasi-phase-matched lithium
  niobate microrings,} {\protect\JournalTitle{Optica}} \textbf{6}, 1244--1245
  (2019).

\bibitem{Wang:waveguide_shg}
C.~Wang, C.~Langrock, A.~Marandi, M.~Jankowski, M.~Zhang, B.~Desiatov, M.~M.
  Fejer, and M.~Lon\v{c}ar, \enquote{Ultrahigh-efficiency wavelength conversion
  in nanophotonic periodically poled lithium niobate waveguides,}
  {\protect\JournalTitle{Optica}} \textbf{5}, 1438--1441 (2018).

\bibitem{Okawachi:20}
Y.~Okawachi, M.~Yu, B.~Desiatov, B.~Y. Kim, T.~Hansson, M.~Lon\v{c}ar, and
  A.~L. Gaeta, \enquote{Chip-based self-referencing using integrated lithium
  niobate waveguides,} {\protect\JournalTitle{Optica}} \textbf{7}, 702--707
  (2020).

\bibitem{ayed_singlephoton}
A.~A. Sayem, R.~Cheng, S.~Wang, and H.~X. Tang,
  \enquote{Lithium-niobate-on-insulator waveguide-integrated superconducting
  nanowire single-photon detectors,} {\protect\JournalTitle{Applied Physics
  Letters}} \textbf{116}, 151102 (2020).

\bibitem{jjlu:1percent}
J.~Lu, M.~Li, C.-L. Zou, A.~A. Sayem, and H.~X. Tang, \enquote{Toward 1\%
  single-photon anharmonicity with periodically poled lithium niobate microring
  resonators,} {\protect\JournalTitle{Optica}} \textbf{7}, 1654--1659 (2020).

\bibitem{rluo:OPG}
R.~Luo, Y.~He, H.~Liang, M.~Li, J.~Ling, and Q.~Lin, \enquote{Optical
  parametric generation in a lithium niobate microring with modal phase
  matching,} {\protect\JournalTitle{Phys. Rev. Applied}} \textbf{11}, 034026
  (2019).

\bibitem{Jankowski:20}
M.~Jankowski, C.~Langrock, B.~Desiatov, A.~Marandi, C.~Wang, M.~Zhang, C.~R.
  Phillips, M.~Lon\v{c}ar, and M.~M. Fejer, \enquote{Ultrabroadband nonlinear
  optics in nanophotonic periodically poled lithium niobate waveguides,}
  {\protect\JournalTitle{Optica}} \textbf{7}, 40--46 (2020).

\bibitem{mYu:two_octave}
M.~Yu, B.~Desiatov, Y.~Okawachi, A.~L. Gaeta, and M.~Lon\v{c}ar,
  \enquote{Coherent two-octave-spanning supercontinuum generation in
  lithium-niobate waveguides,} {\protect\JournalTitle{Opt. Lett.}} \textbf{44},
  1222--1225 (2019).

\bibitem{sWang2020:erbium_ln}
S.~Wang, L.~Yang, R.~Cheng, Y.~Xu, M.~Shen, R.~L. Cone, C.~W. Thiel, and H.~X.
  Tang, \enquote{Incorporation of erbium ions into thin-film lithium niobate
  integrated photonics,} {\protect\JournalTitle{Applied Physics Letters}}
  \textbf{116}, 151103 (2020).

\bibitem{fejer:qpmshg}
M.~M. {Fejer}, G.~A. {Magel}, D.~H. {Jundt}, and R.~L. {Byer},
  \enquote{Quasi-phase-matched second harmonic generation: tuning and
  tolerances,} {\protect\JournalTitle{IEEE Journal of Quantum Electronics}}
  \textbf{28}, 2631--2654 (1992).

\bibitem{Weis1985:summary}
R.~S. Weis and T.~K. Gaylord, \enquote{Lithium niobate: Summary of physical
  properties and crystal structure,} {\protect\JournalTitle{Applied Physics A}}
  \textbf{37}, 191--203 (1985).

\bibitem{xSun:oscillation_dynamics}
X.~Sun, H.~Liang, R.~Luo, W.~C. Jiang, X.-C. Zhang, and Q.~Lin,
  \enquote{Nonlinear optical oscillation dynamics in high-q lithium niobate
  microresonators,} {\protect\JournalTitle{Opt. Express}} \textbf{25},
  13504--13516 (2017).

\bibitem{hJiang17_fastresponse}
H.~Jiang, R.~Luo, H.~Liang, X.~Chen, Y.~Chen, and Q.~Lin, \enquote{Fast
  response of photorefraction in lithium niobate microresonators,}
  {\protect\JournalTitle{Opt. Lett.}} \textbf{42}, 3267--3270 (2017).

\bibitem{jWang16_photothermaleffects}
J.~Wang, B.~Zhu, Z.~Hao, F.~Bo, X.~Wang, F.~Gao, Y.~Li, G.~Zhang, and J.~Xu,
  \enquote{Thermo-optic effects in on-chip lithium niobate microdisk
  resonators,} {\protect\JournalTitle{Opt. Express}} \textbf{24}, 21869--21879
  (2016).

\bibitem{aSavchenkov2006:enhancement_PR}
A.~A. Savchenkov, A.~B. Matsko, D.~Strekalov, V.~S. Ilchenko, and L.~Maleki,
  \enquote{Enhancement of photorefraction in whispering gallery mode
  resonators,} {\protect\JournalTitle{Phys. Rev. B}} \textbf{74}, 245119
  (2006).

\bibitem{Villarroel10_prdamage}
J.~Villarroel, J.~Carnicero, F.~Luedtke, M.~Carrascosa, A.~G.-C. {n}es, J.~M.
  Cabrera, A.~Alcazar, and B.~Ramiro, \enquote{Analysis of photorefractive
  optical damage in lithium niobate: application to planar waveguides,}
  {\protect\JournalTitle{Opt. Express}} \textbf{18}, 20852--20861 (2010).

\bibitem{Yariv96_holographicmemory}
A.~Yariv, S.~S. Orlov, and G.~A. Rakuljic, \enquote{Holographic storage
  dynamics in lithium niobate: theory and experiment,}
  {\protect\JournalTitle{J. Opt. Soc. Am. B}} \textbf{13}, 2513--2523 (1996).

\bibitem{Holzgrafe:20}
J.~Holzgrafe, N.~Sinclair, D.~Zhu, A.~Shams-Ansari, M.~Colangelo, Y.~Hu,
  M.~Zhang, K.~K. Berggren, and M.~Loncar, \enquote{Toward efficient
  microwave-optical transduction using cavity electro-optics in thin-film
  lithium niobate,} in \emph{Conference on Lasers and Electro-Optics,}
  (Optical Society of America, 2020), p. FTh4D.5.

\bibitem{yYang03_opticalstorage}
Y.~Yang, D.~Psaltis, M.~Luennemann, D.~Berben, U.~Hartwig, and K.~Buse,
  \enquote{Photorefractive properties of lithium niobate crystals doped with
  manganese,} {\protect\JournalTitle{J. Opt. Soc. Am. B}} \textbf{20},
  1491--1502 (2003).

\bibitem{mLuennemann03:improvements_of_indexchanges}
M.~Luennemann, U.~Hartwig, and K.~Buse, \enquote{Improvements of sensitivity
  and refractive-index changes in photorefractive iron-doped lithium niobate
  crystals by application of extremely large external electric fields,}
  {\protect\JournalTitle{J. Opt. Soc. Am. B}} \textbf{20}, 1643--1648 (2003).

\bibitem{Carmon2004:thermal_behavior}
T.~Carmon, L.~Yang, and K.~J. Vahala, \enquote{Dynamical thermal behavior and
  thermal self-stability of microcavities,} {\protect\JournalTitle{Opt.
  Express}} \textbf{12}, 4742--4750 (2004).

\bibitem{Surya:18}
J.~B. Surya, X.~Guo, C.-L. Zou, and H.~X. Tang, \enquote{Efficient
  third-harmonic generation in composite aluminum nitride/silicon nitride
  microrings,} {\protect\JournalTitle{Optica}} \textbf{5}, 103--108 (2018).

\bibitem{kukhtarev1976:photorefraction}
N.~V. Kukhtarev, \enquote{Kinetics of hologram recording and erasure in
  electrooptic crystals,} {\protect\JournalTitle{Soviet Technical Physics
  Letters}} \textbf{2}, 348--440 (1976).

\bibitem{Schmidt2008:thermal_effects}
C.~Schmidt, A.~Chipouline, T.~Pertsch, A.~T\"{u}nnermann, O.~Egorov,
  F.~Lederer, and L.~Deych, \enquote{Nonlinear thermal effects in optical
  microspheres at different wavelength sweeping speeds,}
  {\protect\JournalTitle{Opt. Express}} \textbf{16}, 6285--6301 (2008).

\bibitem{Hu2020:all_optical}
X.-X. Hu, J.-Q. Wang, Y.-H. Yang, J.~B. Surya, Y.-L. Zhang, X.-B. Xu, M.~Li,
  C.-H. Dong, G.-C. Guo, H.~X. Tang, and C.-L. Zou, \enquote{All-optical
  thermal control for second-harmonic generation in an integrated microcavity,}
  {\protect\JournalTitle{Opt. Express}} \textbf{28}, 11144--11155 (2020).

\end{thebibliography}

\bibliographyfullrefs{references}


\end{document}